\documentclass[aps,pra,showpacs,twocolumn]{revtex4-1}
\usepackage{graphicx}
\usepackage[usenames]{color}
\usepackage{amssymb}

\newcommand{\fig}[1]{Fig.~\ref{#1}}
\newcommand{\be}[1]{\begin{equation}\label{#1}}
\newcommand{\ee}{\end{equation}}
\begin{document}

\title{Direct versus Delayed pathways in Strong-Field Non-Sequential Double Ionization  }

\author{A. Emmanouilidou$^{1,2}$, J. S. Parker$^{3}$,  L. R. Moore$^{3}$ and K. T. Taylor$^{3}$}
\address{
$^1$ Department of Physics and Astronomy, University College London, Gower Street, London WC1E 6BT, United Kingdom\\
$^2$Chemistry Department, University of Massachusetts at Amherst, Amherst, Massachusetts, 01003, U.S.A.\\
$^3$ DAMTP, Queen's University Belfast, Belfast, BT7 1NN, United Kingdom
}

\begin{abstract}
We report full-dimensionality quantum and classical calculations
of double ionization (DI) of laser-driven helium at 390 nm. Good
agreement is observed. We identify the relative importance
of the two main non-sequential DI pathways, the direct---with an almost simultaneous
ejection of both electrons---and the delayed.  We find that
the delayed pathway prevails at small intensities independently of total
electron energy but at high intensities the direct pathway predominates up to a certain upper-limit in total energy which increases with intensity.  An explanation for this increase with intensity is provided.

\end{abstract}
\date{\today}
\maketitle

Double ionization (DI) of the He atom when driven by strong laser fields serves
as a prototype for exploring correlated electron dynamics in strong fields and
is thus a subject of many studies over the last decade and more, see \cite{taylor2007,Becker2006}.
For large intensities of the laser field the two electrons are stripped
out sequentially---sequential DI (SDI) \cite{lambropoulos1985}. For smaller intensities,
non-sequential DI (NSDI) dominates resulting in the ejection of strongly correlated electron pairs.

Focusing on the range of intensities corresponding to NSDI an accepted mechanism
yielding double ionization is provided by the three-step
model \cite{Corkum93PRL} 1) one electron escapes through the field-lowered Coulomb-barrier, 2) it moves
in the strong infrared laser field and 3) it returns to the core to transfer
energy to the other electron remaining in He$^{+}$.
 Strong support for the three-step model has been provided by both
theory \cite{ruiz2006, haan2008, ParkerJPB2003,Prauzner} and experiment \cite{lafon2001,staudte2007,rudenko2007}.
The transfer of energy in step 3 takes place through two main pathways:
Direct---also referred to as simultaneous ionization, SI; Delayed---also referred to as re-collision-induced excitation with subsequent field ionization, RESI \cite{KopoldPRL2000,FeuersteinPRL2001}---with a delay in ionization of approximately one quarter
of a laser period or more.

Currently we still lack an understanding of how the relative importance of the direct and delayed DI events depends on laser intensity and electron energy.
In the present work, we show that the delayed pathway prevails for small intensities independently of total electron energy but, in contrast, at high intensities the SI pathway predominates up to an upper-limit in total energy---we call this the SI-upper-limit (SIUL).
The SIUL shifts upwards with increasing intensity. We find that accurately accounting for the nuclear interaction is crucial in explaining this upward shift.
Our three-dimensional classical technique fully addresses the Coulomb singularity using regularized coordinates, for details see \cite{EmmanouilidouPRA2008a}. This is a major advantage of our classical technique over others that soften the Coulomb potential. The latter cannot accurately describe DI phenomena related to strong interaction with the nucleus and thus cannot account for most of the detailed findings in the present work.

\begin{figure}[h]

\centerline{\includegraphics[scale=0.25,clip=true]{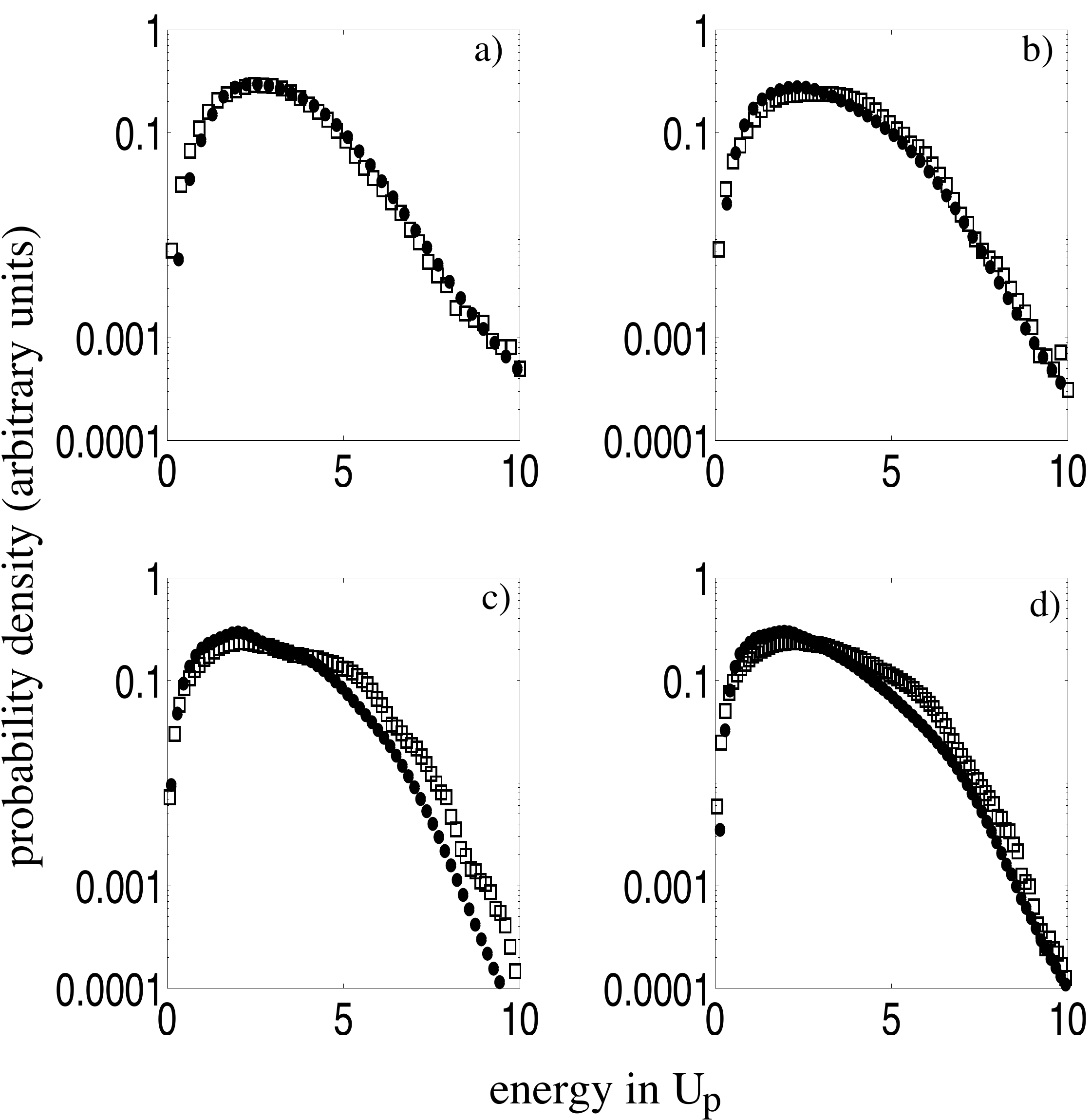}}

\caption{\label{fig:comparison} DI probability density (arbitrary units) for a laser
pulse of 390 nm with intensities: a) 7 $\times 10^{14}$ b) 9 $\times 10^{14}$ c) 13 $\times 10^{14}$
and d) 16 $\times 10^{14}$ W/cm$^2$. The area under each curve is equal to one. The ``smoothed" quantum
results are denoted by black circles, while the classical ones by open black squares. }

\end{figure}

The first finding we report in \fig{fig:comparison}, is surprisingly good agreement over an important range of 390 nm laser
intensities between our classical results and full-dimensionality quantum ones for helium DI energy spectra.  The total energy
in \fig{fig:comparison} is expressed in
units of ponderomotive energy, $U_{p}=E_{0}^2/(4\omega^2)$, a natural choice when comparing different intensities.  The
laser pulse used in the classical calculations is  $E(t) = E_{0}(t) \cos(\omega t)$ and is linearly
polarized along the z-axis. The pulse envelope is defined as $E_{0}(t) = E_{0}$ (a constant) for $0 < t < 6 {\rm T}$
and $E_{0}(t) = E_{0} \cos^{2}(\omega (t-6{\rm T})/12)$ for $6{\rm T} < t < 9{\rm T}$ with T the period of the field.  
The quasiclassical model we use entails one electron
tunneling through the field-lowered-Coulomb potential
with a quantum tunneling rate given by the ADK formula \cite{ADK}. The longitudinal momentum is zero while the
transverse one is given by a Gaussian distribution. The remaining electron is modeled by a microcanonical
distribution \cite{microcanonical}.  An advantage of our classical propagation is that we employ regularized coordinates \cite{regularized} (to
account for the Coulomb singularity) which results in a faster and more stable numerical propagation.
 In
the quantum calculations, for details see \cite{smyth1998}, the pulse has also a 3 field-period ramp-on.  The
laser intensity range considered is important
because at 9 (12) $\times 10^{14}$ W/cm$^2$  the maximum
return energy of the re-colliding electron (3.2 $U_{p}$ within the simplest three-step
model \cite{Corkum93PRL}) equals the first excitation (ionization) energy of ground state
He$^{+}$.

The agreement between classical and quantum results is especially surprising on at least two counts.  Firstly, the
classical approach indeed produces a DI yield at a low laser intensity ($7 \times 10^{14}$ W/cm$^2$), considerably
below the collisional excitation threshold intensity of $9 \times 10^{14}$ W/cm$^2$ at 390 nm.  DI at
such a low laser intensity was long speculated to be either entirely a quantum effect
or only possible by means of repeated re-collisions.
The agreement between classical and quantum in \fig{fig:comparison} clearly negates the former speculation and
analysis of classical DI trajectories below will also negate the latter.  The second surprise is the
agreement over the value of total energy beyond
which the DI probability density falls exponentially---in the following we refer to this energy value as the cut-off.  This energy cut-off value increases
from 5.2 $U_p$ at $7 \times 10^{14}$ W/cm$^2$ to about 7.8 $U_p$ at $13 \times 10^{14}$ W/cm$^2$ in
both calculations. Such increase in the energy cut-off value has previously been reported from the quantum results \cite{parker2006} at this wavelength. The excellent line-up
of the classical with the bench-mark quantum results in \fig{fig:comparison} strongly motivates us to seek deeper
understanding of the DI process through an analysis of the contributing classical trajectories.

We identify the two main DI energy transfer classical trajectory pathways
by using the time delay between the re-collision of the free electron with the parent ion and
the onset of ionization of the second electron \cite{EmmanouilidouPRA2009}. For a definition of the time of ionization see \cite{EmmanouilidouPRA2008a}. In the direct ionization
pathway (SI) both electrons are ionized simultaneously very close to the re-collision time. In
the delayed ionization pathway, the re-colliding electron excites the remaining electron but does not ionize it.  The
electron is subsequently ionized at a peak (RESIa) or at a zero (RESIb) of the laser electric
field \cite{KopoldPRL2000,FeuersteinPRL2001}.  We find that as the intensity increases from 7 $\times 10^{14}$ to 13 $\times 10^{14}$ W/cm$^2$, the contribution of the SI pathway to the total DI yield increases from 25.6\% to 53.4\% while that of the combined RESIa plus RESIb pathway decreases from 62.0\% to 41.8\%. Thus, the SI pathway's contribution to total DI prevails for high intensities.

We now focus on the relative contributions of the direct and delayed pathways to DI within various energy regimes, see Table I.
A convenient way of defining these energy regimes is to use the cut-off energies as boundaries.
 For each intensity, we consider
an upper energy regime bounded from below by the energy cut-off; an intermediate energy regime, if 5.2 $U_{p}$ differs from the cut-off energy, bounded from below by 5.2 $U_{p}$; and finally
a lower energy regime below 5.2 $U_{p}$. As we show below, the effect of the nucleus becomes important at all intensities for energies above 5.2 $U_{p}$, thereby justifying our choice of this boundary value. Note $5.2=2+3.2$, where 3.2 $U_{p}$ is the maximum re-collision energy (3-step model) and 2 $U_{p}$ is the maximum energy a ``free" electron gains from the laser field.

Table \ref{tab:table1} presents the breakdown of the relative contribution of the direct and delayed pathways for three different intensities.  At each intensity, traversing the
respective cut-off leads to a change in the contributions to DI from both the SI and the RESIb pathways.  At
the smallest intensity, 7 $\times 10^{14}$ W/cm$^2$, the totalled RESIa plus RESIb contribution is the major one in
both energy regimes considered---particularly so in the higher energy regime above the cut-off.  At
intensities 9 $\times 10^{14}$ W/cm$^2$ and 13 $\times 10^{14}$ W/cm$^2$, the delayed pathway remains
the predominant one for energies above the cut-off.  However, as the intensity increases
from 7 $\times 10^{14}$ W/cm$^2$ to 13 $\times 10^{14}$ W/cm$^2$, the contribution of the SI
trajectories becomes increasingly more important below the cut-off energy, changing from 27.7 \% to 62.0\%.
 For the higher intensities in Table I it is clear that the SI pathway prevails to an upper-limit in energy---the SI-upper-limit (SIUL)---which increases with intensity.  We show below the crucial role the nucleus plays in this increase in the SIUL.

\begin{table}
\begin{center}
\caption{\label{tab:table1}The \% contributions, in the given energy regimes, to DI
at 7 $\times 10^{14}$ W/cm$^2$, 9 $\times 10^{14}$ W/cm$^2$ and 13 $\times 10^{14}$ W/cm$^2$
from SI, RESIa and RESIb pathways. The energy cut-off has a value of 5.2 $U_p$, 6.5 $U_p$
and 7.8 $U_p$ at these three intensities respectively.}

%\begin{indented}

%\item[]

\begin{tabular}{c|ccc}

\multicolumn{4}{c} { }\\

\multicolumn{4}{c}{}\\

%  & \multicolumn{3}{c} {Energy Regime }\\

 7 $\times 10^{14}$ W/cm$^2$                & SI &  RESIa &RESIb\\

\hline

Below 5.2 $U_{p}$ & 27.7 &  14.3&40.9\\

Above  5.2 $U_{p}$ & 11.4&15.6&64.7\\

\hline

%\hline

\multicolumn{4}{c} { }\\

\multicolumn{4}{c}{}\\

%  & \multicolumn{3}{c} {Energy Regime }\\

9 $\times 10^{14}$ W/cm$^2$                 & SI &  RESIa &RESIb\\

\hline

Below 5.2 $U_{p}$ & 45.1 &  12.0&30.4\\

5.2 $U_{p}$ to 6.5 $U_p$ & 43.9&11.7&39.9\\

above 6.5 $U_p$ & 12.0&13.2&67.8\\

\hline

%\hline

\multicolumn{4}{c} { }\\

\multicolumn{4}{c}{}\\

%  & \multicolumn{3}{c} {Energy Regime }\\

  13 $\times 10^{14}$ W/cm$^2$              & SI &  RESIa &RESIb\\

\hline

Below 5.2 $U_{p}$ & 53.4 &  14.0&24.3\\

5.2 $U_{p}$ to 7.8 $U_p$ &62.0&11.6&21.6\\

above 7.8 $U_p$ & 32.0&15.9&47.0\\

\end{tabular}

%\end{indented}
\end{center}
\end{table}

\begin{figure}[h]
\centerline{\includegraphics[scale=0.35,clip=true]{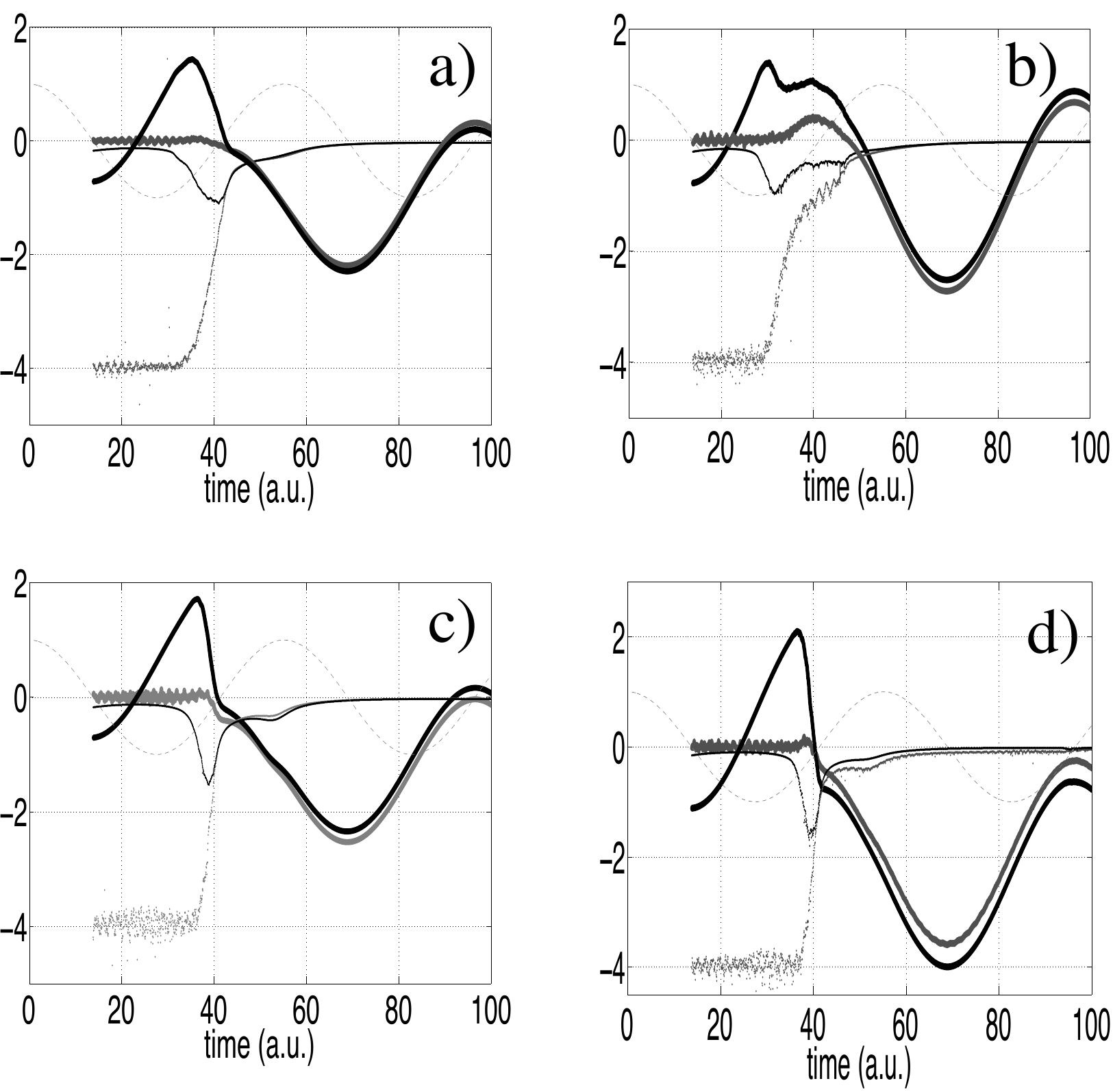}}
\caption{\label{fig:pz1pz2SI} We plot $\left <p_{1,z} \right>$ (thick black line), $\left<p_{2,z}\right>$ (thick grey line), $\left
<Z/r_{1}\right>$ (black dots) and $\left <Z/r_{2}\right >$ (grey dots). These quantities are shown to scale with all coordinate axes
measured in atomic units.  Also shown is the laser field (thin dashed line), which has been arbitrarily scaled so as to be visible on
the plot. Electron 1 is the re-colliding electron, $r_{i}$ is the distance of electron $i$ from the nucleus, and $Z=2$.  The plotted quantities are shown for the SI pathway, for  a) 7 $\times 10^{14}$ W/cm$^2$ at energies below 5.2 $U_p$,  b) 13 $\times 10^{14}$ W/cm$^2$ at energies below 5.2 $U_p$, c) 7 $\times 10^{14}$ W/cm$^2$ at energies above 5.2 $U_p$, and d) 13 $\times 10^{14}$ W/cm$^2$ at energies above 7.8 $U_p$.}
\end{figure}

\begin{figure}[h]
\centerline{\includegraphics[scale=0.35,clip=true]{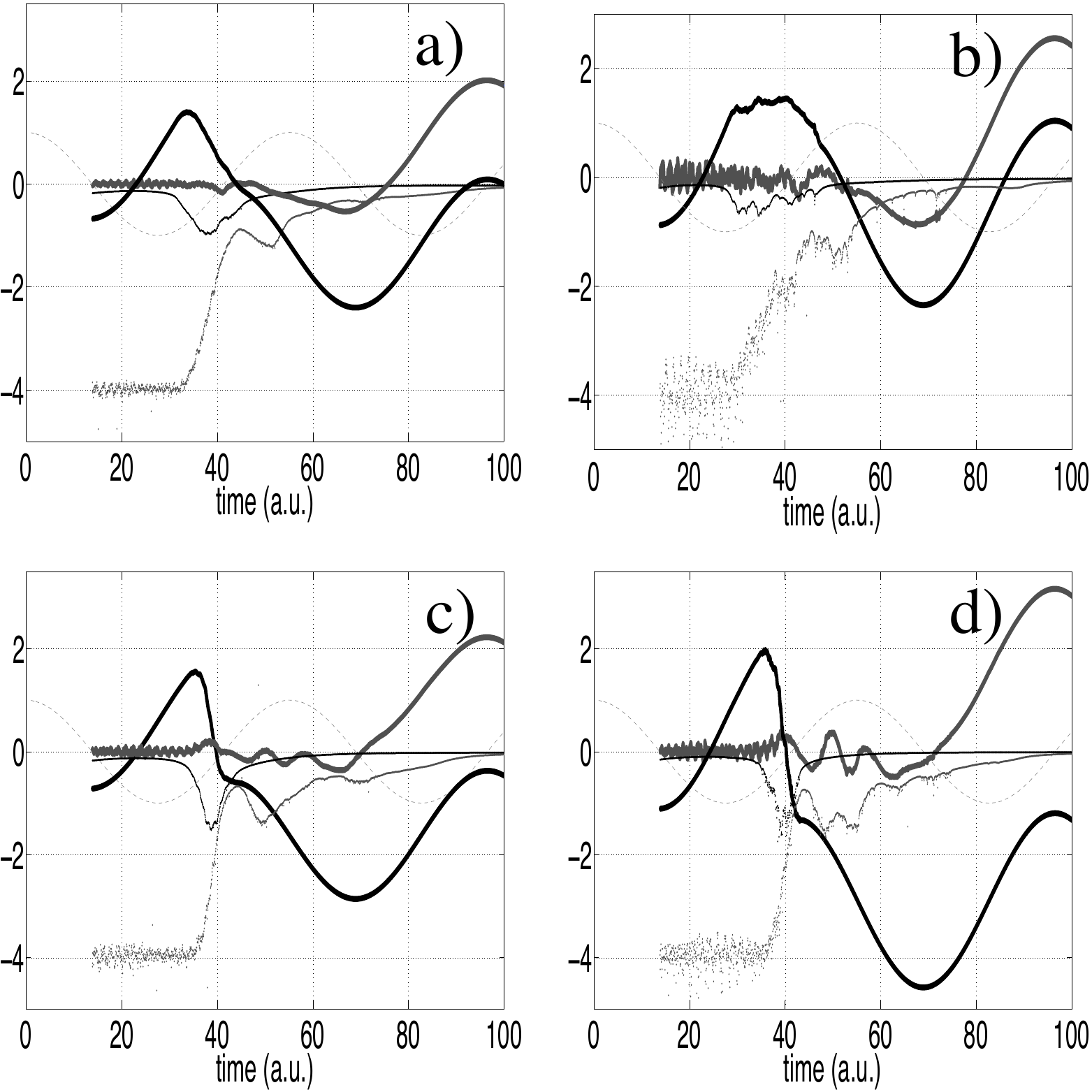} }
\caption{\label{fig:pz1pz2RESI}Same as in \fig{fig:pz1pz2SI} but for the RESIb pathway. }
\end{figure}

We next explore the characteristics and general properties of the DI pathways for some
intensities and energy regimes addressed in Table \ref{tab:table1}. We do so by focusing
on the momentum component of each electron along the polarization axis (the $z$-axis).  We plot
for each electron the average of this component, $\left<p_{1,z}\right>$ and $\left<p_{2,z}\right>$,  for
the SI pathway in \fig{fig:pz1pz2SI} and for the RESIb pathway in \fig{fig:pz1pz2RESI}. The time of re-collision
is the time of minimum distance between the two electrons, identified in \fig{fig:pz1pz2SI} and \fig{fig:pz1pz2RESI} as
the time at which there is a sudden rise/dip of the nuclear contribution to the potential energy of
the second/first electron.  In the SI pathway, electron 2 ionizes at a
time close to the time of re-collision while in RESIb electron 2 ionizes around T/2 later. 
In both \fig{fig:pz1pz2SI} and \fig{fig:pz1pz2RESI} we consider only those trajectories where the re-collision
occurs at the first return of the re-colliding electron to the nucleus. For 390 nm, this is found to be
the most important contribution to the SI and RESI pathways, even for the low intensity
of 7 $\times 10^{14}$ W/cm$^2$. Nevertheless, multiple returns of the re-colliding
electron are explicitly accounted for in Table I. In addition, the general properties of SI and RESIb pathways
for one return of the re-colliding electron, described below, hold true for multiple returns---the only difference
being the re-collision time.

We first address the detailed dynamics at the smallest intensity, 7 $\times 10^{14}$ W/cm$^2$. The time of re-collision
is very close to (2/3)T in accord with a maximum energy
re-collision in the simplest version \cite{Corkum93PRL} of the three-step model. For both energy regimes, the re-colliding electron
in the SI,  \fig{fig:pz1pz2SI} a) and c), and RESIb,
\fig{fig:pz1pz2RESI} a) and c), pathway first loses energy ($\left <p_{1,z} \right>$ suddenly reduces)
to the second electron. It is then pulled by the field---which in the
meantime has changed sign---as well as by the nucleus, in a direction opposite to its incoming direction
before re-collision.   The significant interaction of the re-colliding electron with the nucleus is also
seen as an almost discontinuous change in $\left< p_{1,z}\right>$ shortly after (2/3)T in \fig{fig:pz1pz2SI} a) and c) and
also in \fig{fig:pz1pz2RESI} a) and c). In the SI pathway the field also pulls promptly-ionizing electron 2 in the same
direction as the re-colliding electron,
while for the RESIb pathway electron 2 ionizes later at the next zero of the field. Even though the
interaction of the re-colliding electron with the nucleus is more pronounced for total energies above 5.2 $U_{p}$ the
differences between the two energy regimes are small for both pathways.


At higher intensities the behaviour of $\left<p_{1,z}\right>$ against time differs markedly as we go from
one energy regime to the next, most noticeably so at 13 $\times
10^{14}$ W/cm$^2$. For energies below 5.2 $U_{p}$,  \fig{fig:pz1pz2SI} b) and \fig{fig:pz1pz2RESI} b)
the interaction of the re-colliding electron with the
nucleus is small in both pathways. The re-collision takes place at a time close to T/2
rather than at (2/3)T (the time of maximum energy re-collision in the three-step model). The momentum of the
re-colliding electron first decreases due to transfer of energy to
the second electron---dip in $\left <p_{1,z} \right>$ around T/2 in \fig{fig:pz1pz2SI} b). But at this early
re-collision time the laser field does not undergo a sign change and
so continues to pull electron 1 in the same direction as prior to re-collision. In the SI pathway $\left<p_{2,z}\right>$ has the same
sign as $\left<p_{1,z}\right>$ since ionized electron 2 experiences the same direction of pull from the field as does electron 1.
Thus for high enough intensities within the NSDI regime, SI is similar to a field-free
(e,2e) process for small total energies.  This is in agreement with movie analysis of the two-electron quantum wavepackets \cite{taylor2007}.
For the highest energy regime, see \fig{fig:pz1pz2SI} d) and \fig{fig:pz1pz2RESI} d), the time
of re-collision for both pathways becomes close to (2/3)T. The interaction of the re-colliding electron with the nucleus
is very strong (as indeed is the case both below and above 5.2 $U_{p}$  for 7 $\times 10^{14}$ W/cm$^2$). The re-colliding
electron gets sharply pulled back both by the laser field and the nucleus and reverses in direction, as is the case for small intensities.

To sum up, the SI pathway begins as a small contribution at low intensity but dominates for high intensities
provided the total escape energy remains below the SIUL, see Table 1. Below 5.2 $U_{p}$ the re-collision
time is around T/2 and the interaction with the nucleus is small resulting in comparable
and small escape energies of the two ionizing electrons. For low intensities and any
total energy, as well as high intensity and total energy above the SIUL, the RESI pathway
dominates.  To achieve high-energy DI (beyond 5.2 $U_p$ at any laser intensity)
the re-colliding electron must undergo a strong interaction with the nucleus following a gain of near
maximum energy from the laser field---re-collision time close to (2/3)T. For total energy
beyond 5.2 $U_{p}$ the energy sharing between the two electrons
is quite asymmetric.

We now explore the physical reason for the SI-upper-limit (SIUL) shifting to higher $U_p$ values with increasing intensity, see Table 1.
First, we note that despite this shift, the maximum escape energy of electron 2 remains almost constant at around 2$U_{p}$ \cite{parker2006}.
Second, for energies below the rising SIUL, the SI pathway's contribution
to DI increases from 27.7 \% to 62.0\% with increasing intensity, see Table 1. Given these observations, the question
boils down to: Why does the final escape energy of electron 1 in the SI pathway increase dramatically,
by almost 2.5 $U_{p}$ as the intensity increases from
7 $\times 10^{14}$ W/cm$^2$ to 13 $\times 10^{14}$ W/cm$^2$?

\begin{figure}[h]
\centerline{\includegraphics[scale=0.25,clip=true]{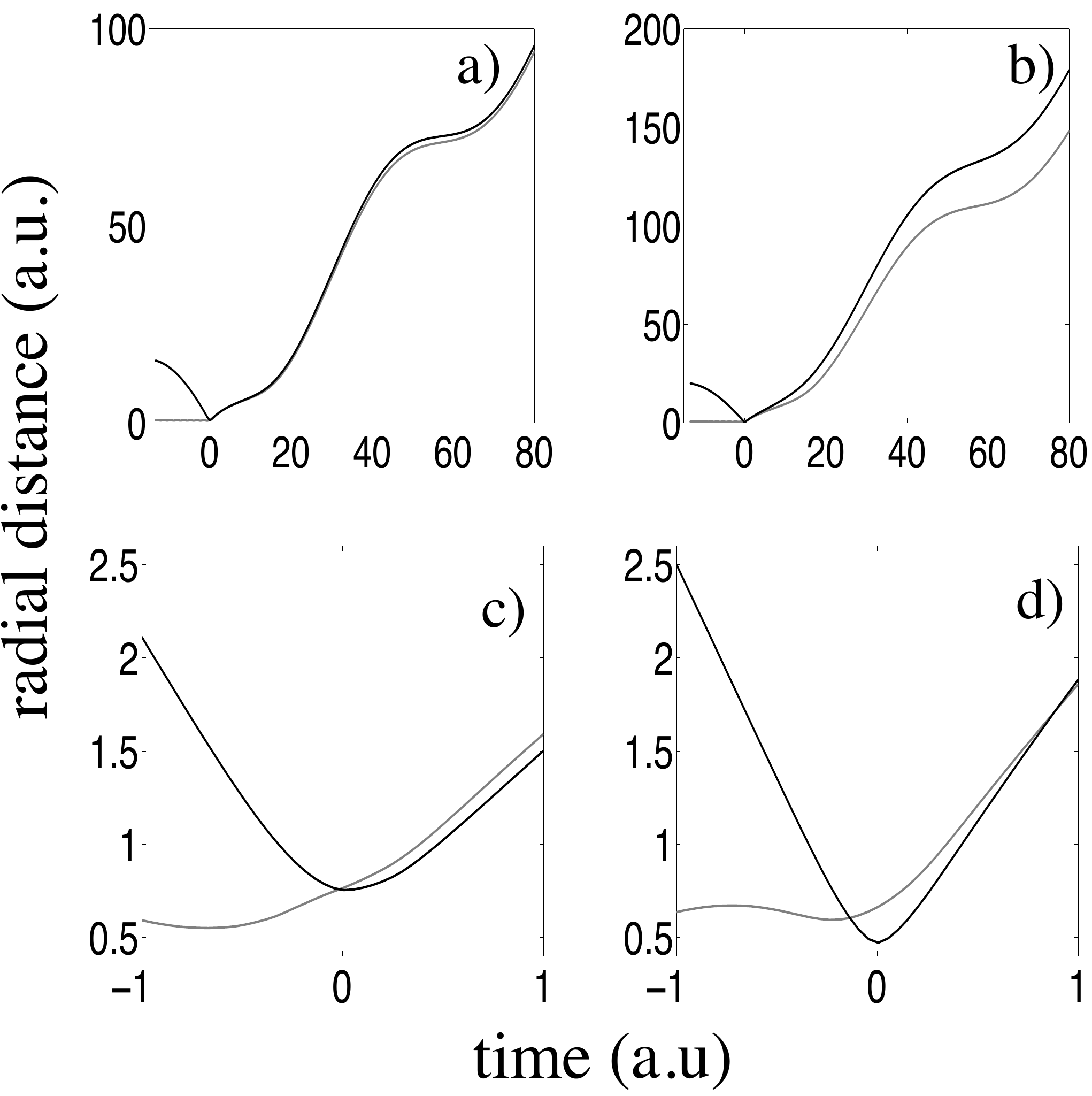}}
\caption{\label{fig:radius} The radial distance of the re-colliding electron (black line) and of the initially bound electron (grey line) for the SI pathway for 7 $\times 10^{14}$ W/cm$^2$ (left) and 13 $\times 10^{14}$ W/cm$^2$ (right) as a function of time for large times (a and b) and small times (c and d). Time zero is the re-collision time.
  }
\end{figure}

For intensities 7 and 13 $\times 10^{14}$ W/cm$^2$ we plot in \fig{fig:radius} the radial distances (from the nucleus) of electron 1 and 2
in the SI pathway taking only total escape energies
extending 1 $U_{p}$ below the intensity-dependent cut-off into account.   The zero of time in these plots corresponds to the re-collision instant.
Frames a) and b) make clear that at  7 $\times 10^{14}$ W/cm$^2$ both electrons escape
with about the same speed but at 13 $\times 10^{14}$ W/cm$^2$, although the electrons escape with faster speeds in line with increasing $U_{p}$,
the re-colliding electron 1 considerably outpaces initially bound electron 2.
Insight on how  electron 1 gains higher escape energy at 13 $\times 10^{14}$ W/cm$^2$ can be gleaned by
examining the bottom frames of \fig{fig:radius} which show magnifications of the top frames at times close to the re-collision instant.
At 7 $\times 10^{14}$ W/cm$^2$, we observe prior to the re-collision that (initially bound) electron 2 has an average
distance from the nucleus of 0.5 a.u.,
as expected for an electron residing in the He$^+$ ground state.
At the instant of re-collision, electron 1 has a radial distance larger than this, at around 0.8 a.u.
In contrast, at 13 $\times 10^{14}$ W/cm$^2$ we see from \fig{fig:radius} d) that at the re-collision instant
electron 1 gets closer to the nucleus than does electron 2.
Thus at 13 $\times 10^{14}$ W/cm$^2$ electron 1 experiences an unscreened nuclear charge making the role of the
nucleus all the greater in controlling this electron and giving it the opportunity to subsequently pick up greater
energy from the laser field.

The increase with intensity of the energy cut-off reported in \cite{parker2006} and of the SI-upper-limit reported here
suggests a connection between the two. It is for future work to establish whether at high intensities the cut-off is a boundary separating energy regimes
where different ionization pathways prevail.

In conclusion  we have found surprisingly good agreement between results from full quantum and classical
calculations for DI energy spectra of 390 nm laser-driven helium over an important range
of laser intensities.  We have analysed the classical results via a unified picture of
DI pathways (direct and delayed) and established how
their relative preponderances change with intensity and total electron escape energy. We find
that the nucleus can play a very important role and that the shift upwards in the SI-upper-limit with
increasing laser intensity comes about through ability of the re-colliding electron to encounter the
unscreened nucleus at higher laser intensity in the direct pathway.  We have shown
that DI at a low intensity (7 $\times 10^{14}$ W/cm$^2$) is possible in a full
classical description where it occurs overwhelmingly via the delayed pathway and strong participation of
the nucleus.  Moreover, an interesting finding of our work is
that at 390 nm, at all intensities explored, the re-collision
at first return of the driven electron dominates all regions of the DI energy spectra.  This is not the case at 800 nm 
 where our preliminary results show collisions other than the first to be the dominant ones. Future work is needed to understand how 
the number of returns of the recolliding electron depends on the frequency of the laser pulse.

This work has been supported by EPSRC and by a Cray Centre of Excellence Award.
A.E gratefully acknowledges funding from EPSRC under Grant No. EPSRC/H0031771, from NSF,
Grant No. NSF/0855403.

\newpage
\bibliographystyle{unsrt}

\end{document}